\documentclass[aps,prl,twocolumn,amsfonts,nobibnotes,superscriptaddress,showpacs]{revtex4-1}

\usepackage{dcolumn}
\usepackage{amsmath}
\usepackage{amssymb}
\usepackage{graphicx}
\usepackage{bm}
\usepackage{color}
\usepackage{braket}
\usepackage[colorlinks=true,citecolor=blue,linkcolor=blue,urlcolor=blue]{hyperref}
\begin{document}

\title{Magnetic field-dependent low-energy magnon dynamics in $\alpha$-RuCl$_3$}

\vspace{2cm} 

\author{Ilkem Ozge Ozel}
	\affiliation{Department of Physics, Massachusetts Institute of Technology, Cambridge, Massachusetts 02139, USA}

\author{Carina A. Belvin}
	\affiliation{Department of Physics, Massachusetts Institute of Technology, Cambridge, Massachusetts 02139, USA}
	
\author{Edoardo Baldini}
\affiliation{Department of Physics, Massachusetts Institute of Technology, Cambridge, Massachusetts 02139, USA}
	
\author{Itamar Kimchi}
\affiliation{Department of Physics, Massachusetts Institute of Technology, Cambridge, Massachusetts 02139, USA}
	
\author{Seunghwan Do}
	\affiliation{Department of Physics, Chung-Ang University, Seoul 06974, Republic of Korea}
	
\author{Kwang-Yong Choi}
	\affiliation{Department of Physics, Chung-Ang University, Seoul 06974, Republic of Korea}

\author{Nuh Gedik}
\email{gedik@mit.edu}
	\affiliation{Department of Physics, Massachusetts Institute of Technology, Cambridge, Massachusetts 02139, USA}

\date{July 31, 2019}

\begin{abstract}
Revealing the spin excitations of complex quantum magnets is key to developing a minimal model that explains the underlying magnetic correlations in the ground state. We investigate the low-energy magnons in $\alpha$-RuCl$_3$ by combining time-domain terahertz spectroscopy under an external magnetic field and model Hamiltonian calculations. We observe two absorption peaks around 2.0 and 2.4 meV, which we attribute to zone-center spin waves. Using linear spin-wave theory with only nearest-neighbor terms of the exchange couplings, we calculate the antiferromagnetic resonance frequencies and reveal their dependence on an external field applied parallel to the nearest-neighbor Ru-Ru bonds. We find that the magnon behavior in an applied magnetic field can be understood only by including an off-diagonal $\Gamma$ exchange term to the minimal Heisenberg-Kitaev model. Such an anisotropic exchange interaction that manifests itself as a result of strong spin-orbit coupling can naturally account for the observed mixing of the modes at higher fields strengths. 
\end{abstract}

\maketitle

\section{I. Introduction}
\begin{figure*}[t]
\begin{center}
\includegraphics[width=2\columnwidth]{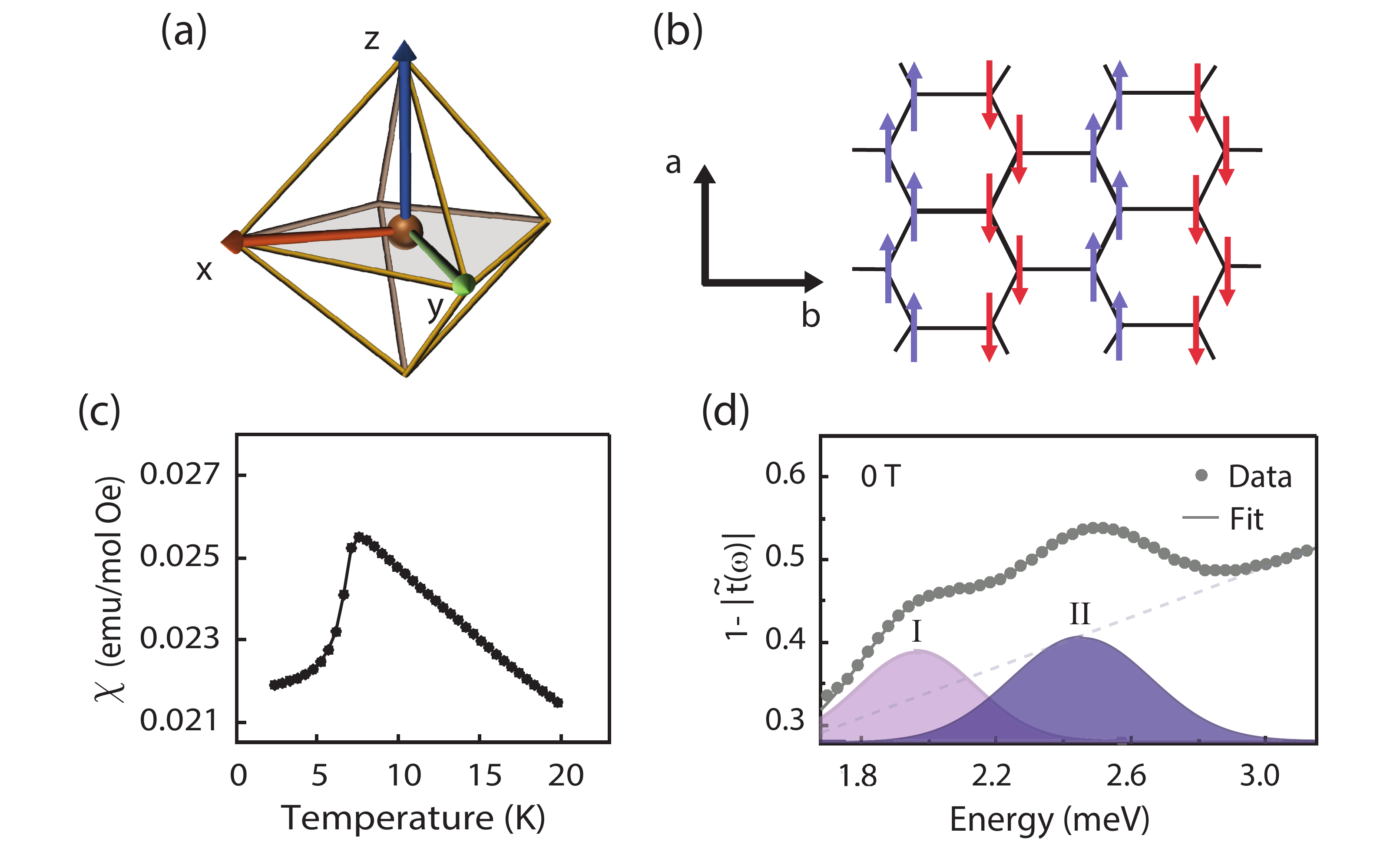}
\caption{(a) Schematic representation of a transition-metal cation (brown sphere) coordinated by six anions (not shown for simplicity) at the vertices of an almost ideal octahedron. This gives rise to Kitaev exchange couplings along the $\hat{x}$, $\hat{y}$, and $\hat{z}$ axes in the Kitaev basis, as shown in red, green, and blue, respectively. (b) Schematic magnetic configuration of zigzag AF order on the 2D honeycomb lattice of $\alpha$-RuCl$_3$ formed by central Ru$^{3+}$ ions below $T\mathrm{_N}$. (c) Temperature dependence of the DC in-plane magnetic susceptibility of $\alpha$-RuCl$_3$ at $H = 1000$ Oe. (d) Representative spectrum of $1-\vert \tilde{t}(\omega)\vert$ as a function of energy below $T\mathrm{_N}$ at 2.1 K measured by TDTS. The spectrum can be modeled phenomenologically by two Gaussian resonances (I and II) plus a linear background (dashed line).}
\label{fig:Figure1}
\end{center}
\end{figure*}

In recent years, considerable interest has been directed towards the realization of unconventional magnetic phases such as the quantum spin liquid (QSL) state \cite{nakatsuji2006metallic, biffin2014unconventional, pilon2013spin, fu2015evidence, norman2016colloquium, witczak2014correlated, shen2016evidence}. Particular focus has been placed on the possible experimental observation of fractionalized quasiparticle excitations in a number of transition-metal compounds with substantial spin-orbit coupling \cite{kobayashi2003structure, kim2008novel, shitade2009quantum, singh2010AF, pesin2010mott, chaloupka2010kitaev, singh2012relevance} following Kitaev's exactly solvable model of anisotropic bond interactions on a two-dimensional (2D) honeycomb lattice \cite{kitaev2006anyons}. In these systems, the transition-metal cations are coordinated by six anions at the vertices of an almost ideal octahedron \cite{kobayashi1992moessbauer, plumb2014alpha}, as illustrated in Fig.~\ref{fig:Figure1}(a), and give rise to spatially dependent exchange interactions \cite{jackeli2009mott, chaloupka2010kitaev, kimchi2011kitaev, kimchi2014kitaev, kimchi2014three, chun2015direct}.

In the quest for the ideal Kitaev material, $\alpha$-RuCl$_3$ has been proposed as a promising candidate. However, unlike ideal QSLs that do not exhibit long-range magnetic order due to strong quantum fluctuations, $\alpha$-RuCl$_3$ enters into a zigzag antiferromagnetic (AF) state below a N\'eel temperature of $T_\mathrm{N}\!\sim$~7~K [Fig.~\ref{fig:Figure1}(b)] \cite{sears2015magnetic, cao2016low}. Nevertheless, spectroscopic probes, including inelastic neutron scattering (INS) \cite{banerjee2017neutron, ran2017spin, do2017majorana, banerjee2018excitations}, spontaneous Raman scattering \cite{sandilands2015scattering, nasu2016fermionic}, time-domain terahertz spectroscopy (TDTS) \cite{little2017antiferromagnetic, wang2017magnetic, shi2018field}, and electron paramagnetic resonance (EPR) \cite{ponomaryov2017unconventional}, have discovered signatures of a field-induced QSL state above 7.5 T in the form of a broad continuum at the 2D magnetic Brillouin zone center. Yet a complete understanding of the origin of these excitations as well as of the spin dynamics is still lacking. Therefore, it is crucial to study the salient features of the spin-wave excitations in the unperturbed or weakly perturbed state.

The zigzag ground state was theoretically shown to be stabilized using the nearest-neighbor Heisenberg-Kitaev (HK) model \cite{chaloupka2010kitaev}, in partial agreement with the experimentally observed magnetic excitation spectrum \cite{banerjee2016proximate}. However, deviations from this spin model were discovered early on, calling for additional terms in the Hamiltonian \cite{singh2012relevance, foyevtsova2013ab, rau2014generic, sears2015magnetic, chaloupka2016magnetic, yadav2016kitaev, janssen2017magnetization, ran2017spin, wang2017theoretical, wolter2017field, suzuki2018effective} such as the off-diagonal $\Gamma$ coupling (a symmetric exchange that is off-diagonal in the Kitaev basis and couples the spin components parallel to the bond orientation) and other terms beyond the nearest-neighbor exchange interactions. Effects of these exchange mechanisms have been observed in the low-temperature magnetization \cite{kubota2015successive}, specific heat \cite{majumder2015anisotropic}, magnetic susceptibility \cite{sears2015magnetic, kubota2015successive, majumder2015anisotropic, lampen2018anisotropic}, and nuclear magnetic resonance spectra \cite{baek2017evidence} of $\alpha$-RuCl$_3$, revealing strong anisotropies for different magnetic field orientations. Despite extensive efforts to explain these observations, to date a definitive consensus on the minimal theoretical model describing the magnetic dynamics in $\alpha$-RuCl$_3$ has not been reached. A promising route to identifying this model is to address the response of the low-energy excitation spectrum to external perturbations \cite{fulde2012correlated}, which directly reflects the complex interplay between different coexisting phases. In this regard, the magnetic field dependence of the magnon modes at terahertz (THz) frequencies in a regime below the threshold for the field-induced QSL state (0 to 5~T) is of particular relevance in $\alpha$-RuCl$_3$.
   
In this study, we combine TDTS with linear spin-wave theory (LSWT) and unveil the behavior of the low-energy magnons in $\alpha$-RuCl$_3$. TDTS is a phase-coherent technique that allows for the direct measurement of complex optical properties in the THz range. Using this approach as a function of external magnetic field, we distinguish features that were previously not resolved by other probes. We observe two magnon modes around 2.0 and 2.4~meV, whose amplitudes and frequencies show a complex field dependence between 0 and 4.8~T. By employing an extended HK model we can capture the zero-field magnon frequencies and the qualitative dependence of the mode frequencies on the applied magnetic field. This allows us to significantly restrict the extensive exchange parameter space that can realize a zigzag ordered state. Our results are suggestive of a scenario in which the off-diagonal $\Gamma$ exchange interaction plays a key role in determining the low-energy physics of the material and imparts a field-induced mixing of modes at higher fields.

This paper is structured as follows: Sec. II describes the experimental methods, Sec. III focuses on the experimental data and the assignment of the observed collective modes, Sec. IV discusses the LSWT analysis, and Sec. V presents the conclusions.

\section{II. Experimental methods}
\subsection{A. Crystal growth and characterization}
The growth of high-quality single crystals of $\alpha$-RuCl$_3$ was carried out using the vacuum sublimation method. Commercial-grade RuCl$_3$ powder (Alfa-Aesar) was dehydrated in a quartz ampoule for a day. The vacuum-sealed ampoule was then placed inside a temperature gradient furnace set at 1080 $^{\circ}$C for 5 h. Next, the furnace was allowed to cool down to 650 $^{\circ}$C at a rate of 2 $^{\circ}$C per hour. The 1:3 (Ru:Cl) stoichiometry of our crystals was confirmed using electron-dispersive x-ray measurements. Our sample was further characterized by magnetic susceptibility measured in an in-plane field of $H = 1000$ Oe, which shows a clear signature of a single magnetic transition at $T\mathrm{_N} \sim$ 7.5 K as determined from the cusp of the curve in Fig.~\ref{fig:Figure1}(c). The appearance of a single sharp magnetic transition at $T_\mathrm{N}$ confirms an ideal AB stacking sequence in the low-temperature phase and a monoclinic $C2/m$ crystalline symmetry at room temperature of our sample, as stacking faults in the form of an ABC-type stacking order have been associated with an additional $T_\mathrm{N}$ of 14 K \cite{kubota2015successive, sears2015magnetic, majumder2015anisotropic, johnson2015monoclinic, reschke2017electronic}. The presence of minimal stacking faults in our sample was also corroborated by single crystal x-ray diffraction. 

\subsection{B. Time-domain terahertz magneto-spectroscopy}
A 5-kHz, 1.55-eV central photon energy, 100-fs Ti:sapphire amplifier system was utilized to generate THz pulses via optical rectification using a ZnTe crystal. The resulting THz radiation was focused onto the sample using off-axis parabolic mirrors, and subsequently detected via electro-optic sampling in a second ZnTe crystal using a weak 1.55 eV gate pulse. For our spectroscopic measurements, we used a home-built THz magneto-optical spectroscopy setup in a transmission geometry. The sample was placed in a helium cryostat with a split-coil superconducting magnet to apply static magnetic fields $H_\mathrm{ext}$ in the 0 to 5~T range at temperatures varying from 2 to 300 K. In our experiments, the sample was zero field cooled, and TDTS was performed in the Voigt geometry. In this measurement scheme, the external magnetic field $H_\mathrm{ext}$ was oriented perpendicular to the THz propagation direction, in the honeycomb plane along the \textbf{b} axis, which is shown in Fig.~\ref{fig:Figure1}(b). The incident THz magnetic field was chosen to lie either along the {\bf{a}} or {\bf{b}} axis. The crystal axes were determined via x-ray diffraction. 

To obtain the transmitted THz field as a function of frequency, the measured time-domain signal was Fourier transformed, yielding a frequency response from 0.4 to 2.5 THz ($\sim$1.65 to 10 meV). For a sufficiently thick sample where temporal windowing of the time-domain signal is appropriate, the frequency-dependent complex transmission coefficient can be calculated by comparing the measured electric field through the RuCl$_3$ sample and a bare aperture reference of the same size,
\begin{equation} 
\tilde{t}(\omega) = \frac{\tilde{E}_\mathrm{sam}(\omega)}{\tilde{E}_\mathrm{ref}(\omega)}\nonumber
= \frac{4\tilde{n}}{(\tilde{n}+1)^2}e^{\frac{i \omega d}{c}(\tilde{n}-1)}. 
\end{equation} 
\par\noindent Here, $\tilde{t}(\omega)$ is the complex transmission coefficient,  $\tilde{E}_\mathrm{sam}$ and $\tilde{E}_\mathrm{ref}$ are the complex frequency-domain THz electric fields of the sample and reference, respectively, $\tilde{n}$ is the complex refractive index of the sample, $\omega$ is the angular frequency, $d$ is the sample thickness, and $c$ is the speed of light in free space. There is no analytical solution to Eq. (1), but $\tilde{n}$ can be numerically extracted following the iterative procedure developed by Duvillaret $et$ $al.$ \cite{duvillaret1996reliable}. The index of $\alpha$-RuCl$_3$ reveals a relatively weak temperature and frequency dependence, and can therefore be assumed to be constant (see Fig.~\ref{fig:FigureS1} of the Supplemental Material \cite{SM}). We obtain $1-|\tilde{t}(\omega)|$ from the magnitude of the complex transmission coefficient.  Owing to the nearly constant index of refraction, this quantity can be simply expressed as a function of the absorption coefficient, 
\begin{equation} 
\label{eq:abs}
|\tilde{t}(\omega)|= \frac{4 n}{(n+1)^2}e^{-\alpha d},\\ 
\end{equation} 
\par\noindent where $\alpha(\omega) = \omega\kappa/c$. This approximation is justified by the relation $n\gg \kappa$, where $\tilde{n} = n-i\kappa$.
\bigskip

\section{III. Experimental results}
\subsection{A. Temperature and magnetic field dependence}
We now focus on the results of our TDTS experiment. Figure~\ref{fig:Figure1}(d) shows a representative spectrum of $1~\!-\!~\vert\tilde{t}(\omega)\vert$ below $T_\mathrm{N}$ with the THz magnetic field \textbf{h} along the crystallographic \textbf{b} direction and no external field. We observe two distinct resonances (labeled I and II) around 2.0 and 2.4~meV, each of which can be described by its amplitude $A$, broadening $\sigma$, and center energy $\Omega$. This allows fitting of the spectra to the following functional form,   
\begin{equation}
f(\omega) = \sum_{i=1}^2 A_i e^{-(\omega-\Omega_i)^2/2{\sigma_i}^2}+ B \omega+ C  
\end{equation}   
\par\noindent in the spectral range from 1.7 to 3.5 meV. In this narrow spectral window, we model the resonances phenomenologically using two Gaussian functions, and the last two terms are used to model the background [dashed line in Fig.~\ref{fig:Figure1}(d)]. The background is found to exhibit a negligible magnetic field dependence.

To clarify the nature of the observed resonances, in the following we study their evolution as a function of temperature $T$ and external magnetic field $H_\mathrm{ext}$. 

\begin{figure*}[t]
\begin{center}
\includegraphics[width=1.7\columnwidth]{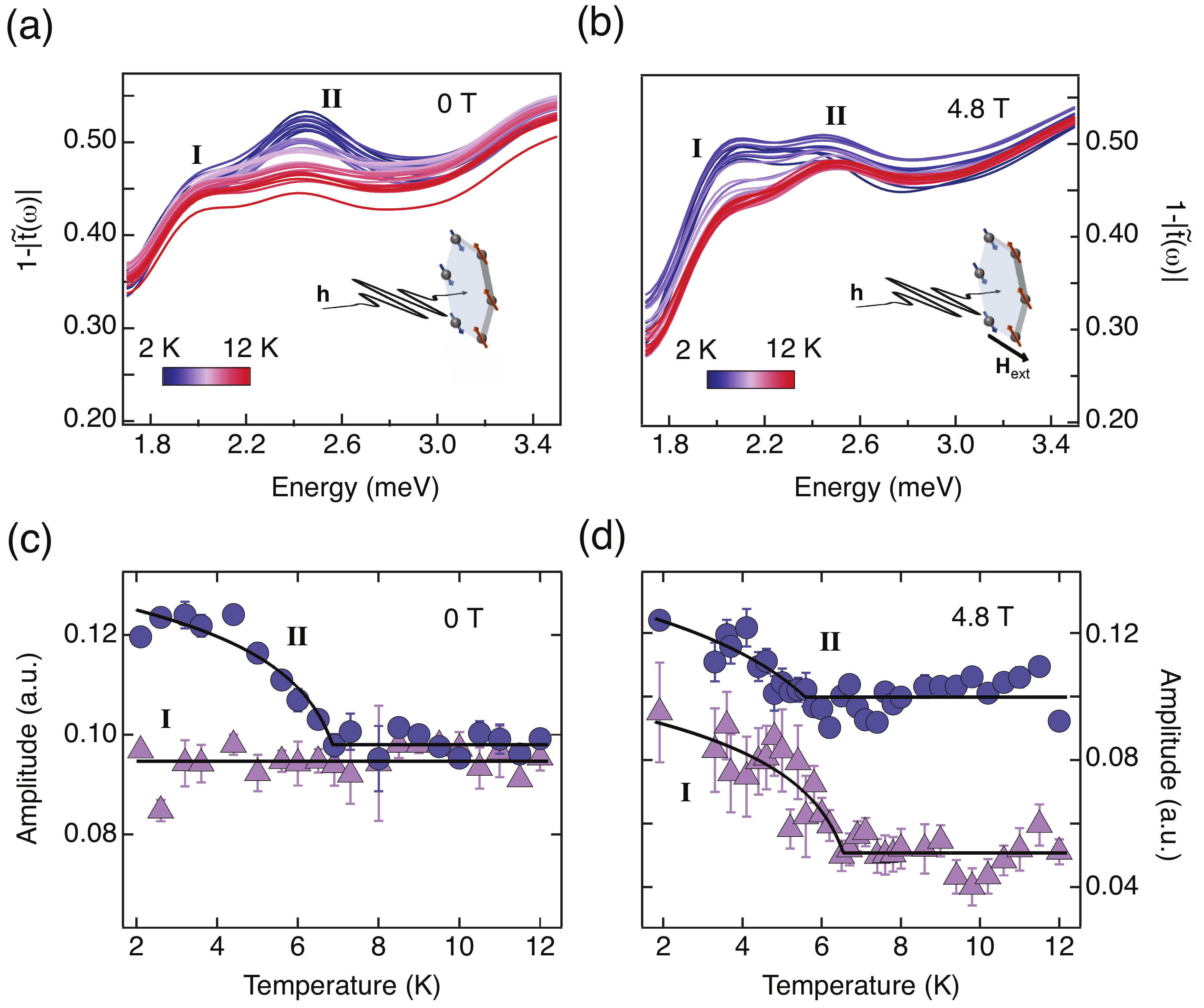}
\caption{THz spectra of $1-\vert \tilde{t}(\omega)\vert$ as a function of temperature at (a) 0~T and (b) 4.8~T, with \textbf{H}$_\mathrm{ext}$,~\textbf{h}~$\parallel$~\textbf{b}. The temperature is varied from 2 to 12 K as indicated by the color bar. Temperature dependence of the amplitudes of modes I (circles) and II (triangles) at (c) 0~T and (d) 4.8~T, obtained by fitting the spectra with two Gaussian profiles and a constant linear background. Error bars indicate the 95\% confidence interval. The solid black lines are guides to the eye.}
\label{fig:Figure2}
\end{center}
\end{figure*}

Figure~\ref{fig:Figure2} compares the temperature dependence of the amplitude of modes I and II at two magnetic field strengths, 0 T [Figs.~\ref{fig:Figure2}(a) and~\ref{fig:Figure2}(c)] and 4.8 T [Figs.~\ref{fig:Figure2}(b) and~\ref{fig:Figure2}(d)]. For $H_\mathrm{ext}$ = 0 and $\bf{h}~\parallel$~\textbf{b}, we observe that the amplitude of resonance II undergoes an order-parameter-like temperature dependence with an onset around $T\mathrm{_N} \sim$~7~K [Fig.~\ref{fig:Figure2}(c), circles]. In contrast, the amplitude of resonance I does not exhibit any discernible temperature dependence [Fig.~\ref{fig:Figure2}(c), triangles]. Strikingly, when a magnetic field of 4.8~T is applied with \textbf{H}$_\mathrm{ext}$~$\parallel$~\textbf{b}, the mode acquires a significant temperature dependence similar to that of resonance II with a critical temperature around 6.5~K [Fig.~\ref{fig:Figure2}(d)]. This onset temperature determined for both resonances matches well with the location of the maximum in the magnetic susceptibility and the specific heat anomaly that was reported previously and was associated with the zigzag magnetic order.

\begin{figure}[h]
\begin{center}
\includegraphics[width=0.93\columnwidth]{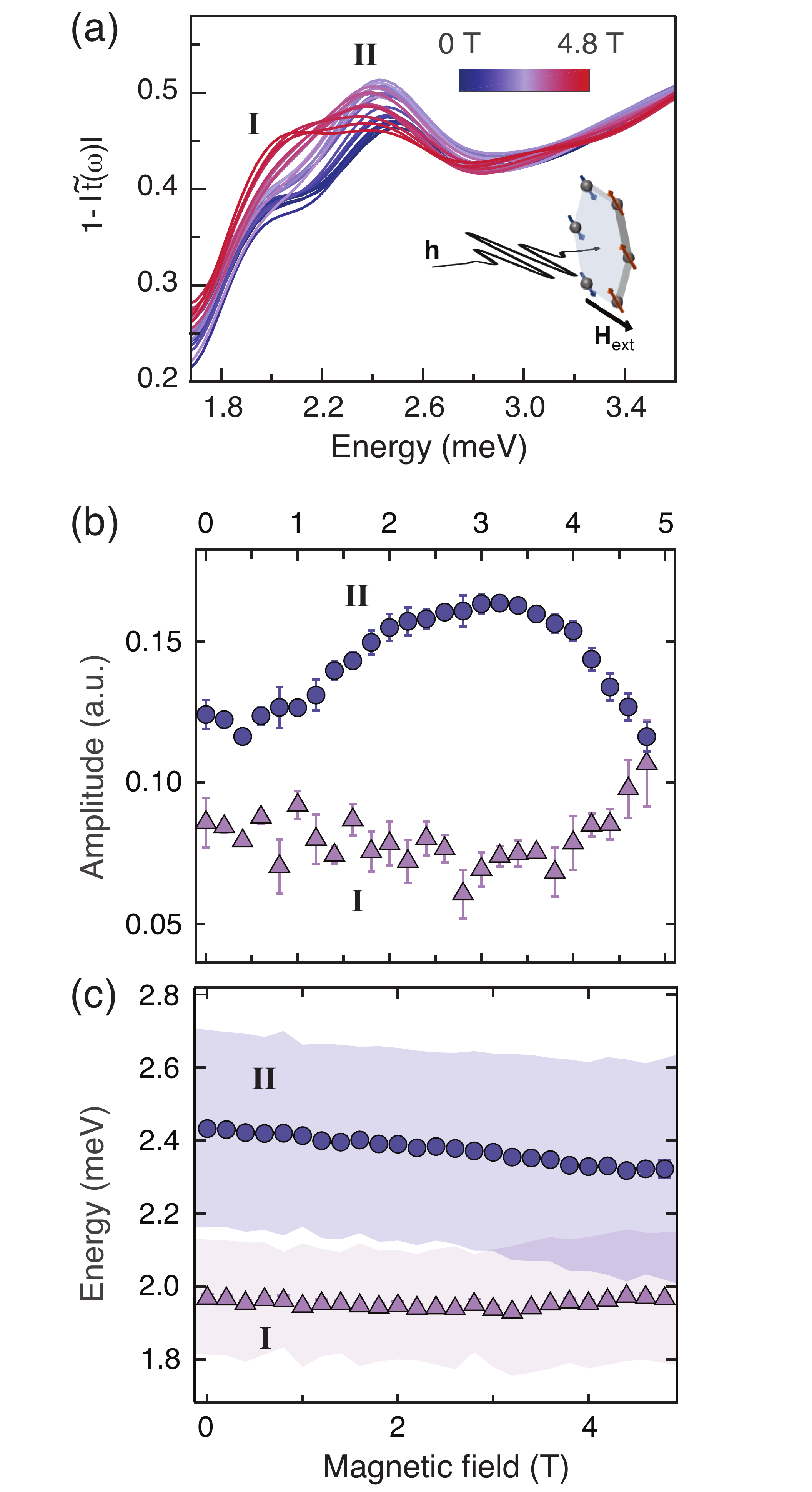}
\caption{(a) THz spectra of $1-\vert \tilde{t}(\omega)\vert$ at 2 K with \textbf{H}$_\mathrm{ext}$,~\textbf{h}~$\parallel$~\textbf{b}. The applied external magnetic field is varied from 0 to 4.8 T as indicated by the color bar. (b) Magnetic field dependence of the amplitudes of modes I (circles) and II (triangles) obtained by fitting the spectra with two Gaussian profiles and a constant linear background. Error bars indicate the 95\% confidence interval. (c) Magnetic field dependence of the energies of modes I (triangles) and II (circles). The lightly shaded areas mark the half-width at half-maximum of the Gaussian line shapes.}
\label{fig:Figure3}
\end{center}
\end{figure}

Next, we study how these resonances evolve as a function of external magnetic field. In Fig.~\ref{fig:Figure3}(a), we compare the spectra taken in the Voigt geometry (\textbf{H}$_\mathrm{ext}$,~\!\textbf{h}~$\parallel$~\textbf{b}, external field varying from 0 to 4.8~T) at 2~K. Figure~\ref{fig:Figure3}(b) tracks the field-dependent amplitude of resonances I and II. Notably, the application of $H_\mathrm{ext}$ first results in an enhancement of resonance II (circles). This initial rise in the mode strength up to 3~T is subsequently followed by a spectral weight redistribution between the two modes at larger fields. Spectra measured for \textbf{h}~$\parallel$~\textbf{a} are presented in Fig.~\ref{fig:FigureS5} of the Supplemental Material \cite{SM}. We note that modes I and II appear in both configurations. While for $T<T_\mathrm{N}$ their relative amplitude depends significantly on the magnitude and direction of \textbf{H}$_\mathrm{ext}$ and \textbf{h}, the spectra do not exhibit a sizable field dependence for $T$~$>~T_\mathrm{N}$ at $T$~=~10~K (see Fig.~\ref{fig:FigureS3} of the Supplemental Material \cite{SM}). 

We also confirmed the existence of two distinct modes in a second $\alpha$-RuCl$_3$ crystal (see Fig.~\ref{fig:FigureS4} of the Supplemental Material \cite{SM}). Although minor differences between samples 1 and 2 are apparent, which can be explained by sample-to-sample variation, overall the spectra exhibit the same features as the field is varied. Similar to what is seen in Fig.~\ref{fig:Figure3}, in Fig.~\ref{fig:FigureS4} mode I also gains notable spectral weight at increasing field strengths. 

\subsection{B. Assignment of the resonances}
The observation of two resonances in the THz spectrum of $\alpha$-RuCl$_3$ suggests that these features can be ascribed to dipole-allowed zone-center collective modes. In order to assign their nature, we consider various possible origins on the basis of the observed behavior. First, we consider phonons. The first-order transition from a monoclinic to a rhombohedral structure that takes place in the temperature range from 60 to 150~K in $\alpha$-RuCl$_3$ \cite{kubota2015successive, glamazda2017relation, reschke2017electronic} has been interpreted as evidence of a magnetoelastic coupling scheme and a natural explanation for the observed phonon anomalies in this material \cite{glamazda2017relation, zhou2018possible}. This raises the question of whether a similar mechanism could explain the unconventional temperature response of mode I, invoking a phonon picture for the observed resonances. However, our THz spectra remain unaltered across this structural transition (see Fig.~\ref{fig:FigureS2} of the Supplemental Material \cite{SM}), suggesting instead a magnetic origin of the modes. Thus, their assignment to back-folded acoustic phonons or to the same magnetic mode split by the presence of occasional stacking faults, which was previously associated with a higher $T_\mathrm{N}$ of 14~K \cite{kubota2015successive, sears2015magnetic, majumder2015anisotropic, johnson2015monoclinic}, can be ruled out by the temperature dependence provided. This observation leads us to conclude that the two resonances are distinct excitations of the underlying zigzag AF order of $\alpha$-RuCl$_3$ with a single $T_\mathrm{N}$ of 7~K. Moreover, the presence of both modes above $T_\mathrm{N}$ (see Fig.~\ref{fig:FigureS2} of the Supplemental Material \cite{SM}), but with smaller amplitude, suggests the persistence of short-range spin correlations in the paramagnetic state above the ordering temperature \cite{do2017majorana, winter2018probing}. 

Consistent with the hypothesis of a magnetic origin of these resonances, we note that mode II was recently observed in independent TDTS \cite{little2017antiferromagnetic, wang2017magnetic, shi2018field, wu2018field} and EPR \cite{ponomaryov2017unconventional} experiments and assigned to a zone-center magnon of the zigzag ordered phase. On the other hand, while signatures of mode I have also been seen in previous measurements \cite{banerjee2018excitations, little2017antiferromagnetic}, this resonance has never been discussed. Specifically, both INS \cite{banerjee2018excitations} and TDTS \cite{little2017antiferromagnetic} spectra taken at different magnetic field amplitudes showed two distinct features at the zone center, similar to ours. While both studies modeled the spectrum in terms of a single spin-wave peak, our extensive temperature and field dependence precludes this interpretation. The field-induced change in the mode response (Fig.~\ref{fig:Figure3}) may result from a modification of selection rules in the magnetic dipole transition matrix elements of strongly spin-orbit coupled $\alpha$-RuCl$_3$, which could potentially also explain the anomalous temperature evolution of mode I at different field strengths that is shown in Fig.~\ref{fig:Figure2}. Although further theoretical studies elucidating the nature of mode I are needed, such changes may emerge from anharmonic effects linked to the symmetry breaking in this material and an associated magnetoelastic coupling below $\sim$150~K \cite{glamazda2017relation, reschke2017electronic, zhou2018possible}. Regardless of their nature, it follows from the markedly different magnetic field dependences of both branches that their assignment as a single mode cannot explain our data. This aspect is of pivotal importance, as the correct identification of the fundamental magnetic excitations places constraints on the exchange interactions governing the spin Hamiltonian, as will be discussed in Sec. IV.

To explain the behavior of the two modes as a function of magnetic field, we note that the threefold rotational symmetry of the $\alpha$-RuCl$_3$ honeycomb layers leads to the appearance of the zigzag order in three distinct domains, related by a spin-orbit-coupled rotation. At zero field, these equivalent domains coexist with ordering wave vectors parallel to the $x$, $ y$, and $z$ bonds ($\vec{Q}_1$, $\vec{Q}_2$, and $\vec{Q}_3$, respectively) \cite{sears2017phase, winter2018probing}. It is expected that the domains do not align along a particular direction in the absence of a field, as the rotational symmetry is preserved. In contrast, in the low-field regime up to 2.5 T, our data reveal clear characteristics of domain rearrangement, in agreement with earlier studies \cite{sears2017phase, banerjee2018excitations}.

Changes in the domain populations can be inferred from the fact that when $H_\mathrm{ext}$ $\neq$ 0, the orientation of local moments across the sample depends on the magnetic field strength through two mechanisms: (i) Within each domain, ``up'' and ``down'' spins cant towards $H_\mathrm{ext}$ through a particular functional form, and (ii) the fraction of spins within each domain varies as a function of $H_\mathrm{ext}$. Classically, it is the fluctuations of these local moments that produce the resonance modes. Ultimately, the system will favor an arrangement of moments that minimizes the exchange energy, which can mainly be achieved when the zigzag chains are oriented perpendicular to the applied field.    

Additional insight and confirmation for the domain-rearrangement scenario were revealed by the dependence of both resonances on an applied field for \textbf{H}$_\mathrm{ext}~\parallel$~\textbf{b} and \textbf{h}~$\parallel$~\textbf{a} (see Fig.~\ref{fig:FigureS5} of the Supplemental Material \cite{SM}). In this configuration, we observe that the amplitude of mode II decreases substantially when $H_\mathrm{ext}~>~$1~T, while mode I remains largely unchanged. This is in stark contrast to the initial rise in amplitude of mode II and the subsequent spectral weight redistribution among modes that is observed for \textbf{h}~$\parallel$~\textbf{b} [see Figs.~\ref{fig:Figure3}(a) and~\ref{fig:Figure3}(b)]. This response is consistent with the argument given above that a rotation of the moments will take place such that the ordering wave vector becomes parallel to the external magnetic field. A continuous increase in the field strength along the \textbf{b} axis will eventually give rise to the preferential selection of the domain with wave vector $\vec{Q}_3$ that is parallel to the \textbf{b} axis (or $z$ bond) in conjunction with a suppressed population of the remaining two domains ($\vec{Q}_1$ and $\vec{Q}_2$) in order to satisfy the exchange interactions that stabilize the AF zigzag order. We find that a complete suppression of these domains occurs around 2~T based on the onset of the plateau region of mode II in Fig.~\ref{fig:Figure3}(b).

\section{IV. Theoretical analysis}
\subsection{A. Minimal spin model}
For our LSWT calculations, we consider the following spin Hamiltonian on a honeycomb lattice
\begin{eqnarray}\nonumber
\hat H&=& \sum_{\braket{ij}} \left[J{\boldsymbol{S}}_i\cdot {\boldsymbol{S}}_j + K S^{\gamma}_i S^{\gamma}_j + \Gamma (S^{\alpha}_i S^{\beta}_j \right. + \left.S^{\beta}_i S^{\alpha}_j)\right]\\
&& - g \mu_\textrm{B} \boldsymbol{H}_\textrm{ext}\cdot \sum_i {\boldsymbol{S}}_i
\end{eqnarray}
\par\noindent where $J$, $K$, and $\Gamma$ represent the Hamiltonian exchange parameters for the Heisenberg, Kitaev, and symmetric off-diagonal $\Gamma$ term, the sum $\braket{ij}$ is over all nearest neighbors, and $g$, $\mu_\textrm{B}$, and $H_\mathrm{ext}$ in the Zeeman term correspond to the $g$ factor, the Bohr magneton, and the external magnetic field, respectively. Here, $\alpha$ and $\beta$ are perpendicular to the Kitaev spin axis $\gamma$. The zigzag order is a collinear order at wave vector $\bf{M}$ in the 2D Brillouin zone. For the Hamiltonian we consider, we find that at zero field the spin moment may be oriented anywhere within the plane through the Bloch sphere that is perpendicular to the ordering wave vector $\bf{Q}$. This relation between real space and the spin Bloch sphere arises from the strong spin-orbit coupling of the Hamiltonian. 

To determine the dispersion of magnetic excitations at finite magnetic fields, we compute the spin-wave spectrum in the partially polarized zigzag AF ordered spin configuration (i.e., the classical ground state at nonzero magnetic fields). Here, the zeroth-order starting point for the spin-wave calculations is a four-sublattice noncollinear magnetic configuration that is a function of $H_\mathrm{ext}$ and the various spin-orbit-coupled magnetic exchanges. For simplicity, we focus on magnetic field orientations that are perpendicular to the plane along which the spins are confined in zero field, i.e., parallel to $\bf{Q}$, and take the ordering wave vector to be only along one type of bond direction, say, $z$ bonds. Canting of the local moments along the field is then a linear process in the field magnitude. We work with magnetic field magnitudes below the saturation field of 7.5~T.

For a given set of values of $H_\mathrm{ext}$ and the Heisenberg, Kitaev, and $\Gamma$ spin exchanges, we first compute the orientation of the zigzag-ordered spins in the classical ground state of the model and then calculate the spectrum of spin fluctuations using standard Holstein-Primakoff substitution within the local spin basis. Consequently, the local polarized moment $m$ (where $m=1$ corresponds to the fully polarized classical state) is found to be $m = 2B[2J+K-\Gamma/2+\sqrt{K^2-K\Gamma +(9/4)\Gamma^2}]^{-1}$. Here, $B$ is the Zeeman term including the $g$ factor and the Bohr magneton. This relation is consistent with that found in \cite{janssen2017magnetization}. We note that the LSWT analysis for such strong spin-orbit coupling was recently compared with exact diagonalization \cite{winter2017breakdown, winter2018probing}, which shows agreement with the dispersion at low energies and additional magnon breakdown effects at higher frequencies.

Throughout this work, we restrict ourselves to a minimal three-parameter model for the exchange couplings including only the nearest-neighbor terms. Due to strong spin-orbit coupling, LSWT is expected to break down. Correspondingly, next-leading-order corrections to the linear spin-wave Hamiltonian would not fully capture the highly nonlinear effects that arise in the real quantum system. While additional higher-order exchange terms have been shown to produce a good description of the spin dynamics (especially further-neighbor Heisenberg interactions) \cite{kimchi2011kitaev,chaloupka2016magnetic,janssen2017magnetization,winter2017breakdown,winter2018probing}, we remark that such corrections are only expected to modify the dispersion away from the zone center. Below we focus only on the two lowest-energy modes, where our spin-wave analysis is expected to be robust. 

The determination of the exchange interaction terms for the spin Hamiltonian is based on two criteria. Our primary focus is on identifying parameter sets that can realize the zigzag state and simultaneously match our experimentally observed magnon resonances at two distinct energies as a function of field at the magnetic zone center. Additional emphasis is given to finding a good correspondence between the calculated magnon dispersion and the spin-wave spectra obtained via inelastic neutron scattering at zero field along the high-symmetry directions. In these earlier studies, gapped spin excitations with minima near 2 meV at the $\bf{M}$ point of the Brillouin zone as well as a local minimum at the zone center were observed \cite{ran2017spin,banerjee2018excitations}. In this respect, we will consider three parameter regimes that stabilize zigzag order in $\alpha$-RuCl$_3$ with zero-field modes close to the experimentally observed energies of 2.0 and 2.4 meV.

\subsection{B. LSWT in a magnetic field}
Irrespective of the detailed microscopic description of the precessional spin motion, our experimental findings suggest that anisotropic exchange mechanisms beyond the pure Kitaev interaction play a dominant role in $\alpha$-RuCl$_3$, consistent with previous works \cite{sears2015magnetic, winter2017breakdown, janssen2017magnetization, ran2017spin, suzuki2018effective, lampen2018anisotropic}. To provide a quantitative estimate of these couplings, a clear observable is the evolution of the spin-wave energies with an external field, as this quantity can be captured within the framework of LSWT. The experimentally determined energies as functions of field for resonances I and II are shown in Fig.~\ref{fig:Figure3}(c) with purple and blue symbols, respectively. Mode I possesses only a weak field dependence, shifting slightly towards higher energies as the field increases, whereas mode II softens more steeply with an applied field. 

\begin{figure*}[t]
\begin{center}
\includegraphics[width=2\columnwidth]{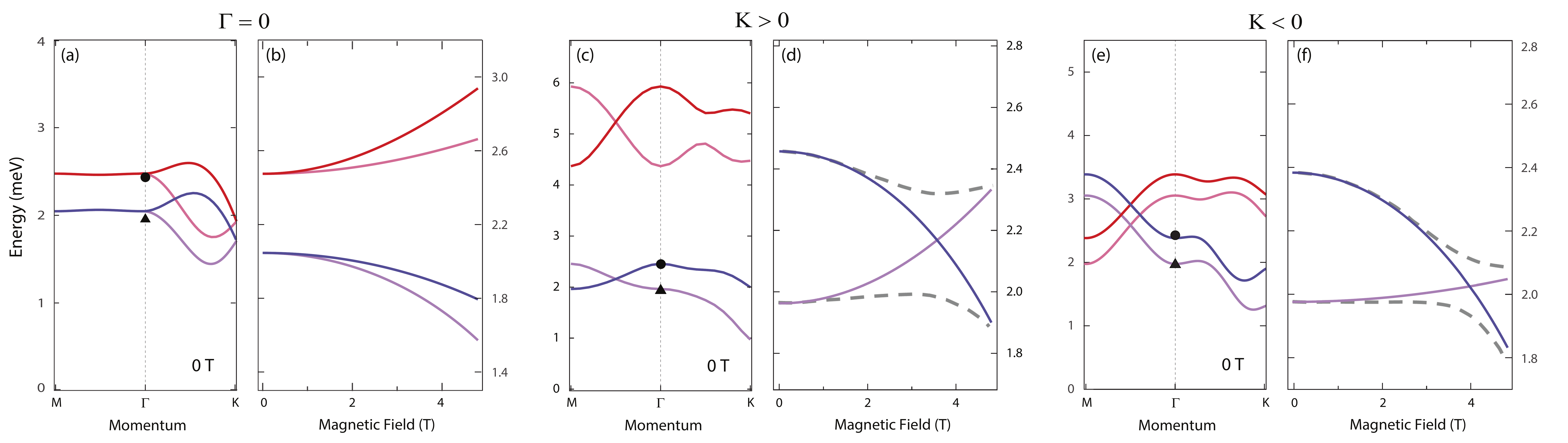}
\caption{(a), (c), and (e) Magnon energy-momentum dispersion relation obtained from LSWT for $H_\mathrm{ext}$~=~0 along high-symmetry directions of the magnetic Brillouin zone and (b), (d), and (f) energy versus field of the relevant lowest two magnon branches in $\alpha$-RuCl$_3$ at the zone center for \textbf{H}$_\mathrm{ext}~\parallel$~\textbf{b} using an (a) and (b) HK model, (c) and (d) HK$\Gamma$ model with $K>0$, and (e) and (f) HK$\Gamma$ model with $K<0$. Dashed lines are guides to the eye indicating the mixing of modes, and the solid symbols mark the experimental points obtained via TDTS.}
\label{fig:Figure4}
\end{center}
\end{figure*}

Next, for a field applied in the \textbf{b} direction, we obtain the magnon dispersions using LSWT in Fig.~\ref{fig:Figure4}. Figures~\ref{fig:Figure4}(a),~\ref{fig:Figure4}(c), and~\ref{fig:Figure4}(e) show the calculated dispersions for $H_\mathrm{ext}$~=~0 along high-symmetry directions of the magnetic Brillouin zone, while Figs.~\ref{fig:Figure4}(b),~\ref{fig:Figure4}(d), and~\ref{fig:Figure4}(f) correspond to the magnetic field evolution of the two lowest-lying magnon branches at the Brillouin zone center. By varying the magnitude of $H_\mathrm{ext}$, we study how the spin-wave energies renormalize under the influence of the magnetic field. We investigate in detail the behavior of the spin waves employing a model Hamiltonian with (1) $\Gamma$ = 0 and finite $J$, $K$ [Figs.~\ref{fig:Figure4}(a) and~\ref{fig:Figure4}(b)], (2) ferromagnetic $J$ ($J<0$), AF $K$ ($K>0$), and $\Gamma >0$ [Figs.~\ref{fig:Figure4}(c) and~\ref{fig:Figure4}(d)], and (3) AF $J$ ($J>0$), ferromagnetic $K$ ($K<0$), and $\Gamma >0$ [Figs.~\ref{fig:Figure4}(e) and~\ref{fig:Figure4}(f)].

As a starting point, it is reasonable to consider a simple model that comprises the least number of exchange terms. It has been pointed out that a $K\Gamma$ description alone is not sufficient to stabilize zigzag order \cite{janssen2017magnetization}. Thus, we explored the regime of finite $J$ and $K$ ($\Gamma$~=~0), with our primary focus being good agreement between spin-wave calculations and the lowest two magnon modes observed at 2.0 and 2.4 meV via TDTS at zero field. We restrict our parameter range to $(J,K)=(-1.75,3.1)$. Although $(1.75,-3.1)$ yields the same zero-field mode energies, here, we do not consider this parameter regime as a zigzag state has been found to exist only in the nearest-neighbor HK model when the Kitaev coupling is AF, i.e., $K>0$. In Fig.~\ref{fig:Figure4}(a), we plot the magnon dispersion at zero field along the high-symmetry directions of the magnetic Brillouin zone. In this coupling scheme, the magnetic order is established via the ferromagnetic (FM) Heisenberg exchange within the chains, while adjacent zigzag chains couple antiferromagnetically through $K>0$. Notably, at zero field, the calculated magnon energies at the Brillouin zone center capture the experimental data points of Fig.~\ref{fig:Figure3}(c) [marked by solid black symbols in Fig.~\ref{fig:Figure4}(a)]. 

We next turn to the field dependence of the calculated magnon dispersions and compare these with our data. The disagreement between the calculated spin-wave dispersion in an applied field [Fig.~\ref{fig:Figure4}(b)], in which the lowest two modes soften, whereas the higher ones bend upward, and the experimental data shown in Fig.~\ref{fig:Figure3}(c) illustrates that the contribution of an off-diagonal $\Gamma$ interaction beyond the nearest-neighbor $J$ and $K$ exchange couplings is crucial. A notable discrepancy is also apparent between the spin-wave spectra obtained by inelastic neutron scattering revealing a noticeable dip at the $\bf{M}$ point \cite{ran2017spin, banerjee2018excitations} and the calculated magnon dispersions in the HK model. Additionally, a significant $\Gamma$ coupling has been suggested to account for the different Curie-Weiss temperatures that were measured for external fields applied parallel and perpendicular to the honeycomb planes \cite{sears2015magnetic, janssen2017magnetization, lampen2018anisotropic}. Below, we demonstrate that a spin model supplemented with a significant anisotropic $\Gamma$ interaction is indeed in better agreement with the experimentally observed magnon behavior in this study. We will further demonstrate that although an FM Kitaev term in our model may potentially explain the empirical field dependence of the modes, our careful search of the parameter space suggests that an AF Kitaev interaction is better at fitting the zone-center spin waves. 

Figure~\ref{fig:Figure4}(c) shows the calculated energy-momentum dispersion relation of four magnon branches at $H_\mathrm{ext}$ = 0 for a dominant $\Gamma$ and a sizable AF Kitaev term. An excellent match is obtained when $J=-0.95$~meV, $K=1.15$~meV, and $\Gamma=3.8$~meV near the Brillouin zone center. This is highlighted by the solid circle and triangle, which denote the values of the magnon energies extracted from our TDTS data for $H_\mathrm{ext}$ = 0. Importantly, a finite $\Gamma$ term is required to reproduce the measured magnetic field evolution of the spin-wave excitations at the magnetic zone center by our TDTS measurements [Fig.~\ref{fig:Figure3}(c)], in addition to the reported gap of $\sim$2~meV seen near the \textbf{M} point in previous neutron scattering studies \cite{banerjee2016proximate, ran2017spin, banerjee2017neutron, banerjee2018excitations}. Qualitative agreement with our experimental results is retrieved, in that resonance I blueshifts with increasing field while resonance II redshifts. The fitted parameters predict a crossing of the two distinct modes at $\sim$3.6~T [Fig.~\ref{fig:Figure4}(d)]. Conversely, our experimental finding points towards the existence of an apparent avoided crossing. Hence, we argue that the correct interpretation of our data presented in Fig.~\ref{fig:Figure3}(c) is a field-induced mixing between the two magnon modes.

To motivate this interpretation, we rely on phenomena arising in other systems that show clear mixing behavior. In general, two energetically close elementary excitations can be considered coupled quantum oscillators when they are characterized by similar energies, the same momentum, and the same symmetry \cite{fleury1968soft}. When the frequencies are brought sufficiently close to each other upon tuning an external parameter ($H_\mathrm{ext}$ in our case), the underlying interaction between the two modes leads to their hybridization, and the mode eigenvectors become indistinguishable. Clear signatures of mode mixing are represented by similar temperature dependences, intermode transfers of spectral weight, and mode frequency repulsion \cite{fleury1968soft, heilmann1981one, iliev2006distortion, osano1982theory}. 

In this respect, the peculiar temperature dependence shown by the amplitude of mode I at 4.8~T (Fig.~\ref{fig:Figure2}(b)) in our experiments, as well as the redistribution of spectral weight occurring between the two modes starting around 3.5~T [Fig.~\ref{fig:Figure3}(b)], is strongly reminiscent of a similar mode-mixing character. By the same token, the two resonances become comparable in amplitude near 4.8~T [see Fig.~\ref{fig:Figure3}(b)], pointing towards an enhanced coupling between the two excitations. This coupling scheme is further supported by the noticeable spin-wave broadening and the concomitant growth of the overlapping region at higher fields (3.0 to 4.8~T), which is bounded by the lightly shaded areas that mark the half-width at half-maximum of the Gaussian line shapes [Fig.~\ref{fig:Figure3}(c)]. 

From previous studies of magnon-magnon interactions, it is known that highly nonlinear effects are large and unavoidable for a strongly spin-orbit coupled Hamiltonian. The off-diagonal anisotropic $\Gamma$ term in particular has been demonstrated to play an important role in nonlinear spin dynamics, giving rise to the breakdown of the single-particle formalism \cite{chaloupka2016magnetic,janssen2017magnetization,winter2017breakdown,winter2018probing}. These effects have, in fact, been highlighted in exact diagonalization calculations \cite{winter2016challenges, winter2017breakdown, winter2018probing} and various other approximation schemes \cite{kim2015kitaev, yadav2016kitaev, winter2016challenges,janssen2017magnetization}, in which strong anharmonicity and decay into lower-energy magnons necessarily arise as a consequence of the Kitaev and $\Gamma$ terms in the Hamiltonian. Therefore, it may be anticipated that a considerable mixing between the two spin-wave branches in Fig.~\ref{fig:Figure4}(d) occurs in line with our empirical observation [Fig.~\ref{fig:Figure3}(c)]. 

With such anharmonic effects observed in $\alpha$-RuCl$_3$, a natural question that arises is the relevance of magnetoelastic interactions that have been reported to prevail in this system in the temperature range of $\sim$60$-$150~K \cite{glamazda2017relation, reschke2017electronic, zhou2018possible}. Although there is no direct evidence of a change in the crystal structure in the low-temperature regime near 7~K where the zigzag order is stabilized, it remains to be explored whether and to what extent the strong spin-lattice interactions as revealed by Raman studies and the magnon mixing behavior reported in our current work are related to one another. Such anharmonic magnon interactions are expected since the off-diagonal $\Gamma$ interaction is known to originate from the symmetry breaking of the crystal structure due to lattice distortions \cite{suzuki2018effective}. However, further theoretical and experimental studies are required to investigate the relevance of these effects in the context of the low-temperature behavior of zigzag-ordered $\alpha$-RuCl$_3$.    
 
Last, we demonstrate that our data can also be fitted reasonably well with an alternative set of exchange parameters, in which the Kitaev term is ferromagnetic. This scenario was investigated by several $ab$ $initio$ \cite{yadav2016kitaev,winter2016challenges,kim2016crystal} and experimental \cite{ran2017spin, do2017majorana, koitzsch2017low} studies. The magnon dispersions from our model with dominant ferromagnetic $K$, where $K=-3.50$ meV, $\Gamma=2.35$ meV, and $J=0.46$ meV, are depicted in Figs.~\ref{fig:Figure4}(e) and~\ref{fig:Figure4}(f). We note that our measurements together with LSWT presented herein cannot establish the actual sign of the Kitaev term, i.e., $K<0$ or $K>0$. Nevertheless, our key focus in this study is on highlighting the important role played by the anisotropic $\Gamma$ term in the spin Hamiltonian \cite{winter2017breakdown, suzuki2018effective}, which is confirmed by both parameter sets. Moreover, the identification of two closely spaced spin-wave excitations via TDTS and their respective field evolution allows us to significantly restrict the parameter space to a very narrow window and determine the hierarchy of exchange terms in this spin-orbit coupled material.

\section{V. Conclusions} 
To conclude, we studied the low-energy magnon dynamics of $\alpha$-RuCl$_3$ using time-domain terahertz spectroscopy. Our data suggest the presence of two magnon modes, whose amplitudes and energies as a function of external magnetic field evolve distinctly. From the magnetic field dependence of the magnon energies at the Brillouin zone center and the observed anticrossing behavior near 4.8~T, we infer a set of exchange parameters using linear spin-wave calculations. Our experiments strongly suggest the ubiquity of other exchange mechanisms beyond the simple Heisenberg-Kitaev model, in particular the off-diagonal $\Gamma$ coupling, as well as the importance of nonlinear magnon processes in the spectroscopic signatures of $\alpha$-RuCl$_3$.\\
 
\section{Acknowledgments}
We thank P. A. Lee and Z. Alpichshev for useful discussions. Work at MIT was supported by the U.S. Department of Energy, Basic Energy Sciences, Division of Materials Science and Engineering, Award No. DE-FG02-08ER46521. C.A.B. and E.B acknowledge additional support from the National Science Foundation Graduate Research Fellowship under Grant No. 1122374 and the Swiss National Science Foundation under Fellowships No. P2ELP2-172290 and No. P400P2-183842, respectively. I.K. was supported by the MIT Pappalardo fellowship program. S.D. and K.-Y.C. acknowledge support by the Korea Research Foundation (KRF) Grant No. 2017R1A2B3012642 funded by the Korean government (MEST).\\

\clearpage
\begin{center}
\textbf{\large Supplemental Material}
\end{center}
\renewcommand{\thefigure}{S\arabic{figure}} 
\setcounter{figure}{0} 

\section{I. Index of refraction of $\alpha$-RuCl$_3$}

\begin{figure}[h]
\begin{center}
\includegraphics[width=\columnwidth]{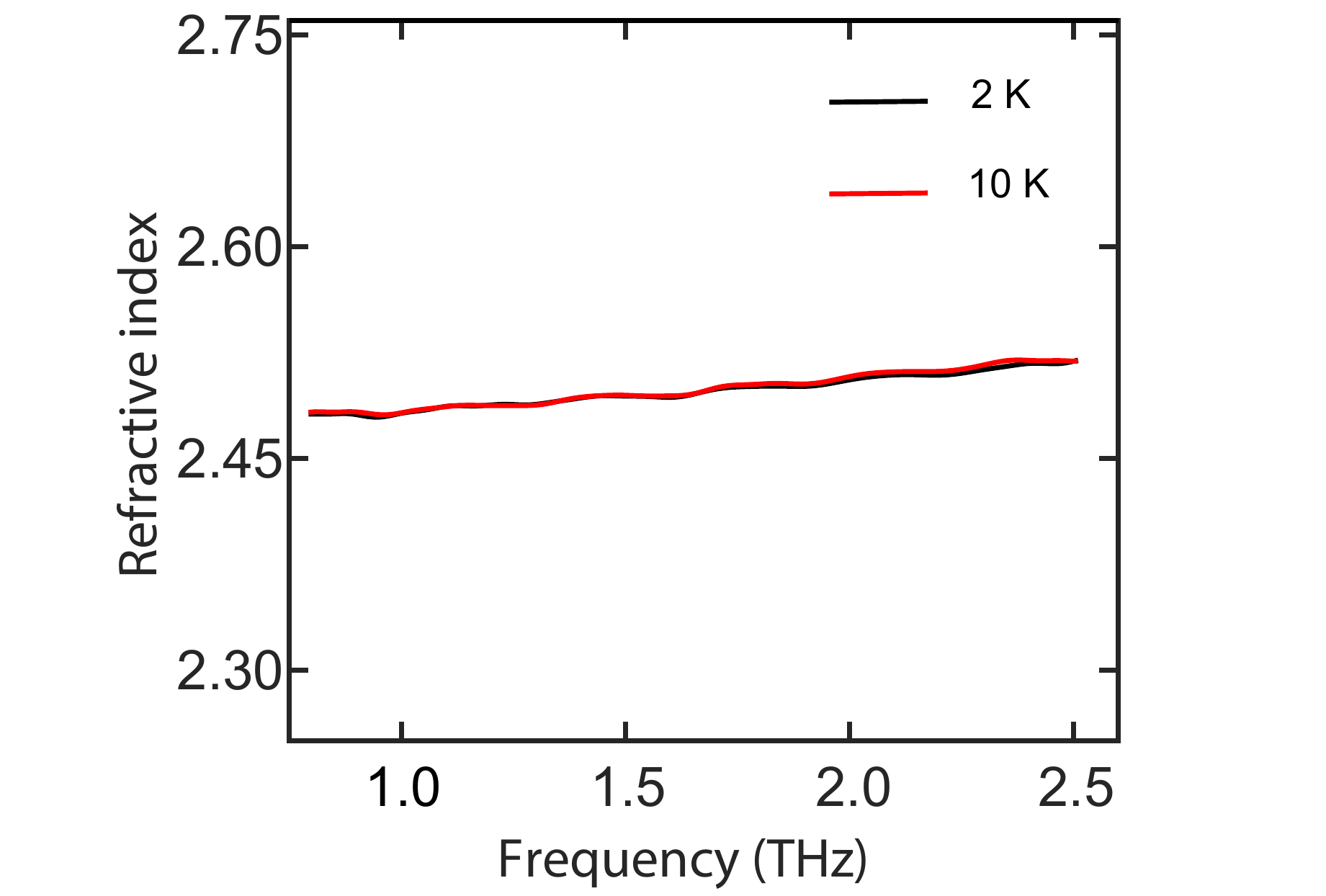}
\caption{The real part of the refractive index of $\alpha$-RuCl$_3$ as a function of frequency at 2~K (black) and 10~K (red).}
\label{fig:FigureS1}
\end{center}
\end{figure}

Figure~\ref{fig:FigureS1} shows the real part of the refractive index $n$ of sample 1, which has a thickness of 0.83~mm, at 2~K (black) and 10~K (red). Only a slight variation in the index (2.48 to 2.52 over the entire spectral range) is observed. There is also no discernible change in the index below and above $T_\mathrm{N}$. 

\section{II. Temperature and field dependence of spectra above $T_\mathrm{N}$}

\begin{figure}[t]
\begin{center}
\includegraphics[width=\columnwidth]{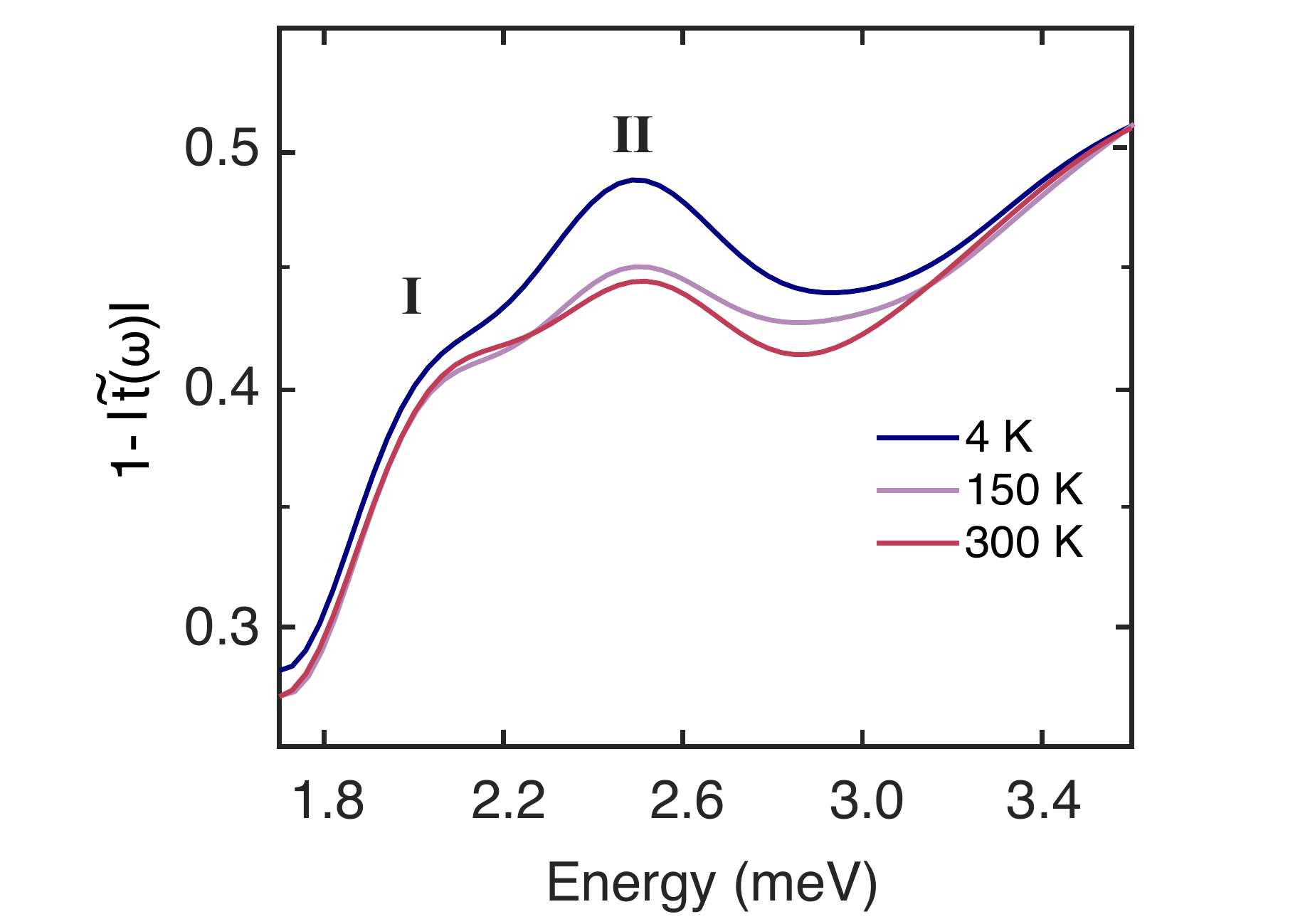}
\caption{Comparison of THz spectra measured at various temperatures up to 300 K.}
\label{fig:FigureS2}
\end{center}
\end{figure}

\begin{figure}[t]
\begin{center}
\includegraphics[width=\columnwidth]{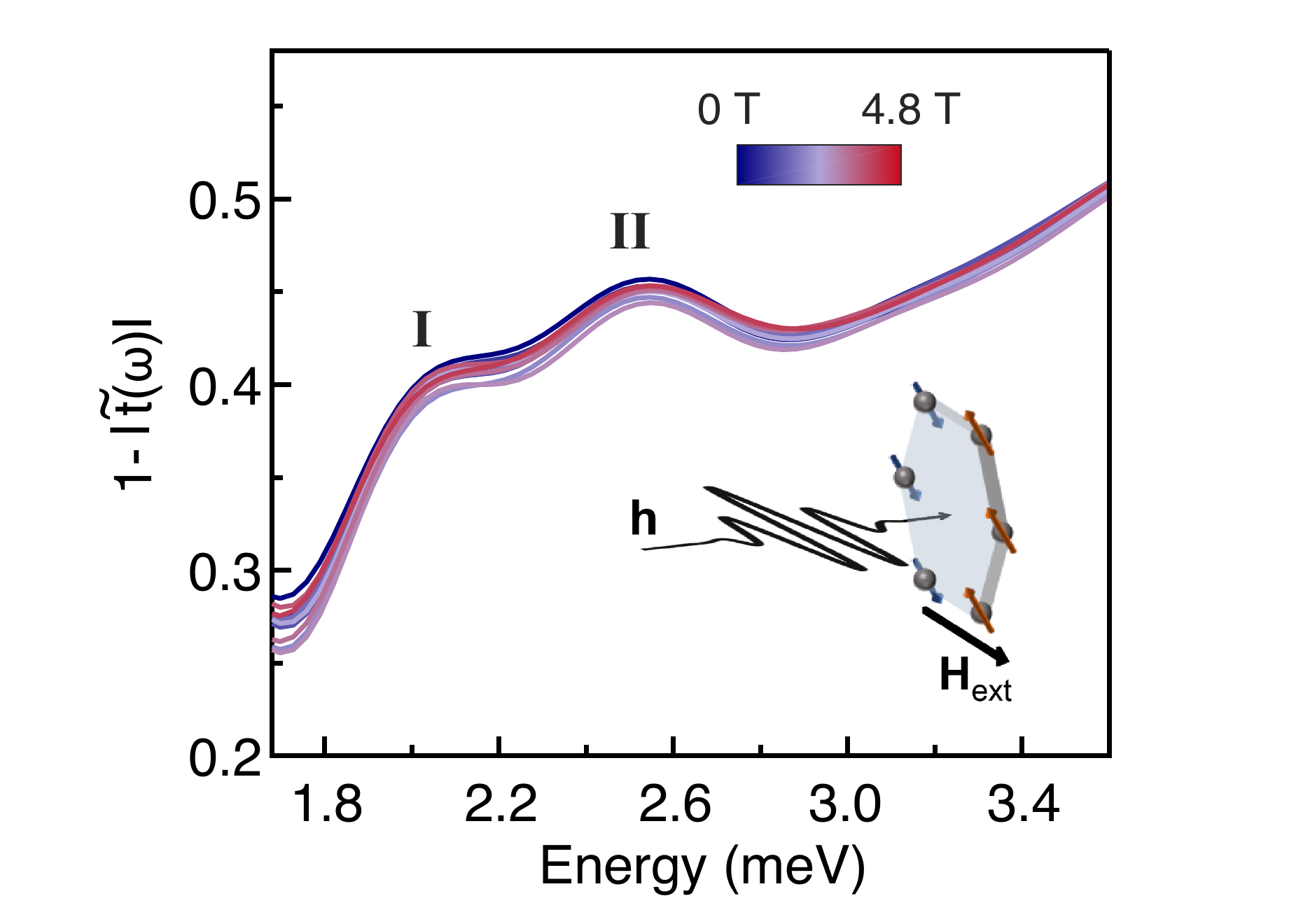}
\caption{THz spectra at 10~K with \textbf{H}$_\mathrm{ext}$,~\textbf{h}~$\parallel$~\textbf{b}. The applied external magnetic field is varied from 0 to 4.8 T as indicated by the color bar.}
\label{fig:FigureS3}
\end{center}
\end{figure}

In our analysis, we do not employ a time-windowing or referencing procedure to get rid of spurious etalon effects and spectral features associated with the high temperature phase in contrast to previous experimental works \cite{little2017antiferromagnetic, wang2017magnetic}. This in turn allows us to spectrally resolve two closely spaced resonances and study their intrinsic behavior as a function of our external tuning parameters. Figure~\ref{fig:FigureS2} shows spectra measured above $T_\mathrm{N}$ in comparison with the spectrum at 4~K. The resonance has a weak residual amplitude at a temperature as high as 300 K, which can be attributed to short-range magnetic correlations that have been reported to survive in the paramagnetic state well above $T_\mathrm{N}$ \cite{do2017majorana, suzuki2018effective, winter2018probing}. Furthermore, it is interesting to note that our data shows no indication of the first-order structural transition that takes place in the 60$-$150~K temperature range \cite{kubota2015successive, glamazda2017relation, reschke2017electronic, park1609emergence}. This can be clearly seen by comparing the features of the spectra above and below this range. The traces, which consist of a double-peak structure sitting on top of a continuum background, remain fairly unchanged except for an increase in the amplitude of mode II below $T_\mathrm{N}$, as shown in Fig. ~\ref{fig:Figure2}(a) of the main text.
 
Additional evidence for the magnon interpretation of these modes, other than the observed order-parameter-like temperature behavior that onsets at $T_\mathrm{N}$ (Fig.~\ref{fig:Figure2}), comes from the absence of a field dependence of the spectra above $T_\mathrm{N}$ at $T$~=~10~K, as given in Fig.~\ref{fig:FigureS3}. The insensitivity of the spectra to the applied magnetic field also reflects that magnetoelastic coupling interactions associated with the structural phase transition from the high-temperature monoclinic to the low-temperature rhombohedral phase are less important in this temperature range of interest. These considerations together with lacking evidence of a structural distortion near 7~K confirm that a magnon picture is most appropriate to explain the observed resonances, although further theoretical and experimental studies are needed to understand the effects of anharmonic spin-phonon interactions in  $\alpha$-RuCl$_3$.  

\section{III. Field dependence of spectra for sample 2}
    
\begin{figure}[t]
\begin{center}
\includegraphics[width=\columnwidth]{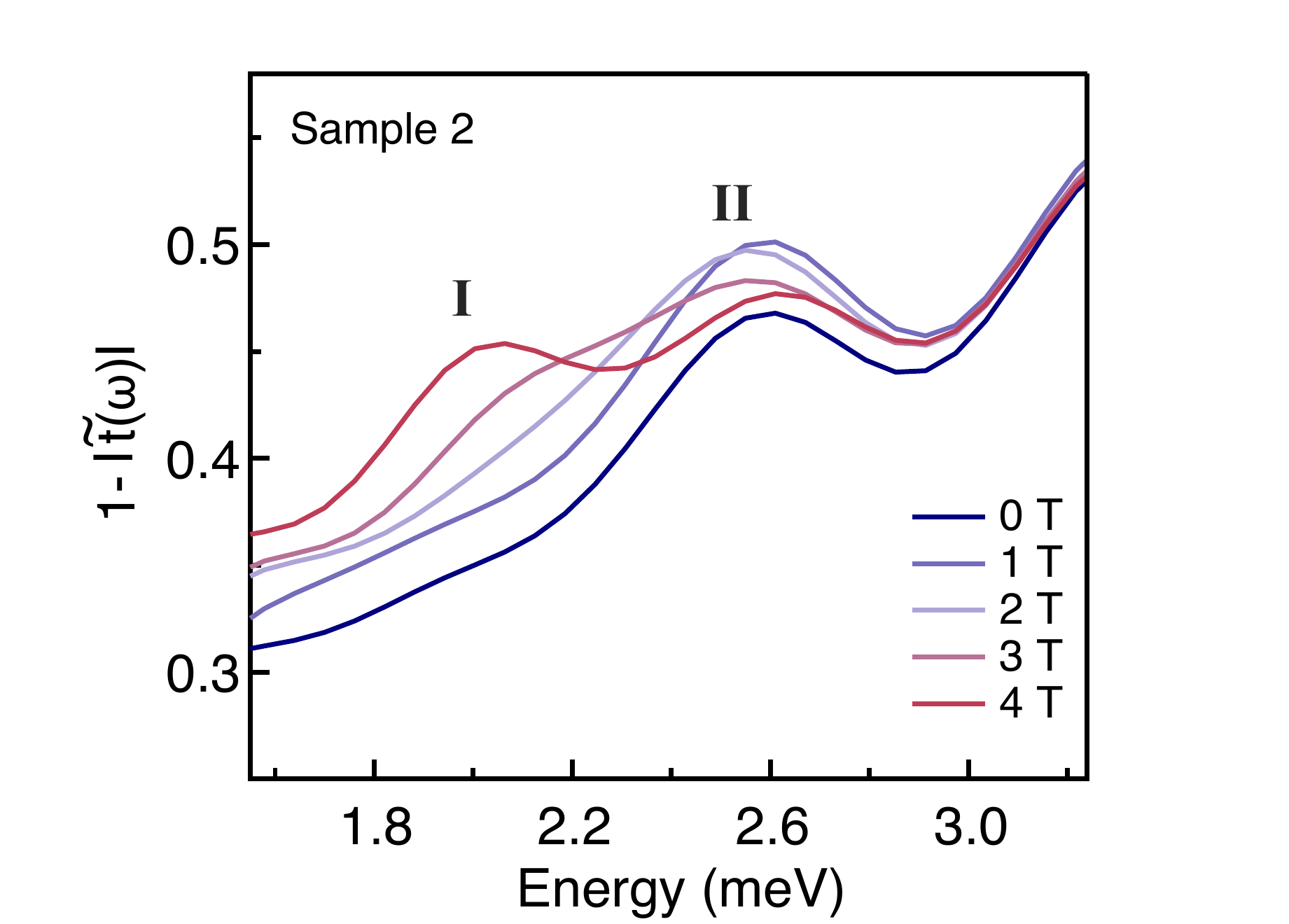}
\caption{THz spectra of sample 2 at 2~K with \textbf{H}$_\mathrm{ext}$,~\textbf{h}~$\parallel$~\textbf{b}.}
\label{fig:FigureS4}
\end{center}
\end{figure}

The existence of two distinct modes was also confirmed in a second $\alpha$-RuCl$_3$ crystal (Fig.~\ref{fig:FigureS4}). While spectra of samples 1 and 2 exhibit minor differences, which can be explained by variations from sample to sample, the field-dependent curves retain the same spectral features to a large extent. In fact, mode I can be distinguished more clearly in sample 2. The robust observation of this mode in a different sample confirms the necessity of a revised spin model of $\alpha$-RuCl$_3$ that can properly capture the low-energy excitations and shed further light into the nature of the observed continuum in this material.      

\section{IV. Polarization dependence of spectra}

\begin{figure}[t]
\begin{center}
\includegraphics[width=\columnwidth]{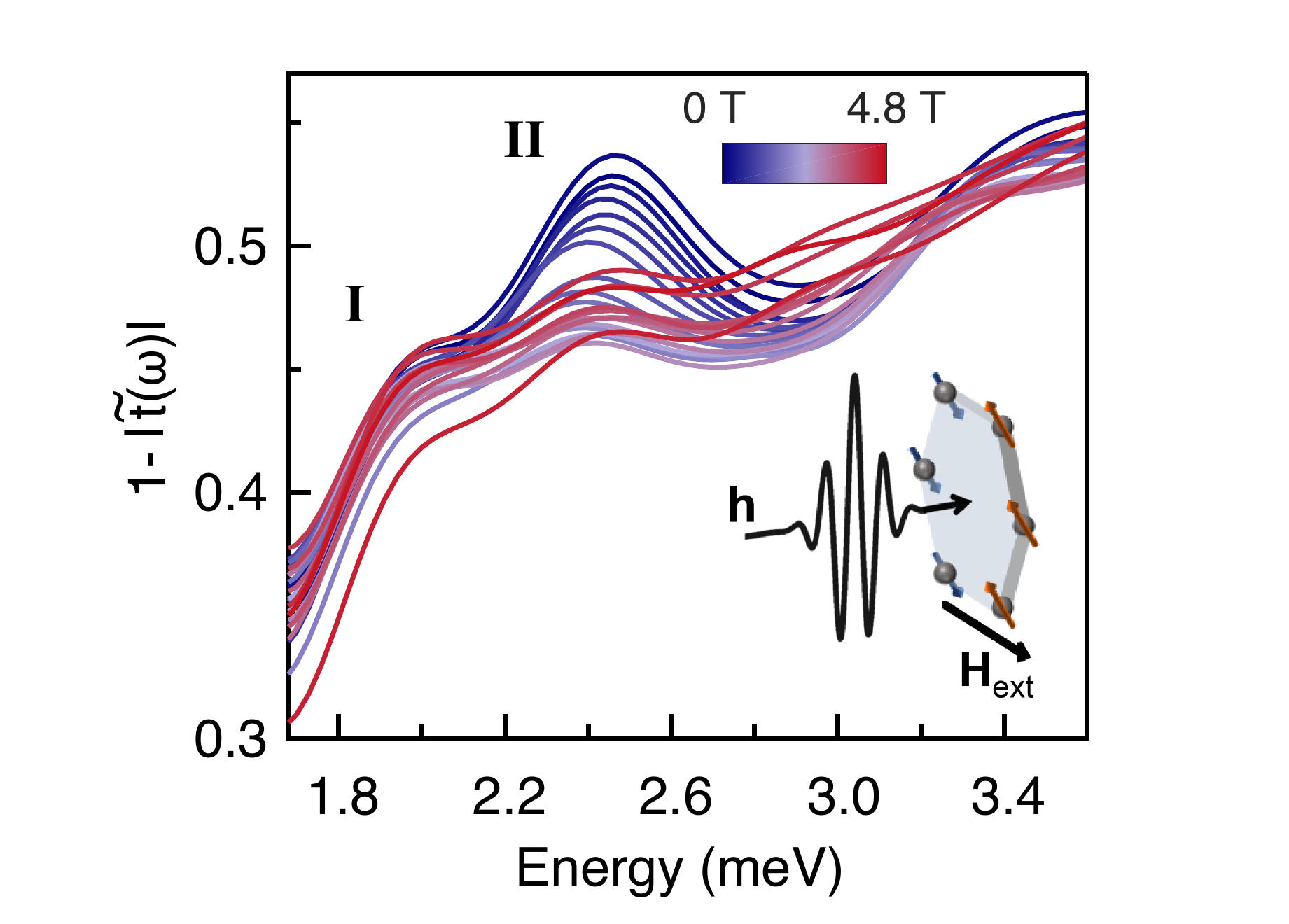}
\caption{THz spectra at 2~K with \textbf{h}~$\parallel$~\textbf{a} and \textbf{H}$_\mathrm{ext}$~$\parallel$~\textbf{b}. The applied external magnetic field is varied from 0 to 4.8 T as indicated by the color bar.}
\label{fig:FigureS5}
\end{center}
\end{figure}

To gain more systematic insight into the behavior of modes I and II, we study how the resonances respond to an applied field for \textbf{H}$_\mathrm{ext}~\parallel$~\textbf{b} and \textbf{h}~$\parallel$~\textbf{a} (Fig.~\ref{fig:FigureS5}). In this configuration, we observe that the amplitude of mode II decreases substantially with $H_\mathrm{ext}$ while mode I remains largely unchanged. This behavior is in stark contrast to the initial rise in amplitude of mode II and subsequent spectral weight redistribution between the modes that was observed for \textbf{h}~$\parallel$~\textbf{b} (see Fig.~\ref{fig:Figure3}). We also note that the amplitude of modes I and II are comparable for fields below 1~T in both field configurations. Above 1~T, we differentiate between the behavior of mode II for \textbf{h}~$\parallel$~\textbf{b} and \textbf{h}~$\parallel$~\textbf{a}. This result agrees with a recent magnetic neutron diffraction study \cite{sears2017phase} in which the complete suppression of a magnetic domain was demonstrated when the ferromagnetic zigzag chains run along the applied field direction for fields as low as 2~T.  

At zero field, the threefold symmetry of the lattice in the 2D limit allows for the coexistence of the zigzag order in three equivalent domains with ordering wavevectors parallel to the $x$, $y$, and $z$ bonds \cite{sears2017phase, winter2018probing}. When an external magnetic field is applied, the spins will align perpendicular to the field direction to minimize the free energy and simultaneously satisfy the exchange interactions that give rise to the zigzag order. However, due to the broken spin rotation symmetry, this can only be achieved when the applied field is parallel to the Ru-Ru bonds (as it is in our case, when \textbf{H}$_\mathrm{ext}~\parallel$~\textbf{b}). Consequently, even at fields as small as 1~T a reorientation of domains will occur, resulting in an enhancement of the domain population with ordering wavevector that is parallel to the \textbf{b} axis (or $z$ bond) in conjuction with a depopulation of the remaining two domains. We find that a complete suppression of these domains occurs around 2~T based on the onset of the plateau region of mode II in Fig.~\ref{fig:Figure3}(b), which is consistent with recent neutron diffraction data \cite{sears2017phase, banerjee2018excitations}. 

More generally, our observations for mode II in both configurations can be understood in terms of moments oscillating along the bond direction and interacting with the magnetic field component $\textbf{h}$ of the THz field that is parallel to the moments \cite{mukhin2004antiferromagnetic}. Since there is no preferrential domain at zero field, it is expected to see comparable amplitudes for both probing configurations of $\textbf{h}$ (along the cystallographic {\bf{a}} and {\bf{b}} directions). At finite fields, however, the spins cant toward the applied field, hence resulting in an increase of the amplitude of mode II when $\textbf{h}~\parallel~\textbf{b}$ (Fig.~\ref{fig:Figure3}) and a decrease when $\textbf{h}~\parallel~\textbf{a}$ (Fig.~\ref{fig:FigureS5}). More generally, at larger fields in the range of 3 to 4.8~T, the complicated dependence of the amplitude of modes I and II may be qualitatively understood as arising from matrix element effects under strong spin-orbit coupling accompanied by a mixing of modes as discussed in the main text. With strong spin-orbit coupling, these elements are generally nonvanishing and depend in a complicated manner on the mode wavefunction and fluctuation directions. 

\bibliographystyle{apsrev4-1}

\end{document}